\documentclass[12pt]{JHEP}
\usepackage{epsfig,amssymb} 

\def\be{ \begin{equation}}          \def\ee{ \end{equation}}
\def\ba{ \begin{eqnarray}}          \def\ea{ \end{eqnarray}}

\def\nn{\nonumber}                  
\def\ap{\alpha '}

  \def\R{\mathbb{R}}
\def\T{\mathbb{T}}
\def\o{\otimes}                     

\def\cedille#1{\setbox0=\hbox{#1}\ifdim\ht0=1ex \accent'30 #1%
 \else{\ooalign{\hidewidth\char'30\hidewidth\crcr\unbox0}}\fi}
\def\gaw{Gaw\cedille edzki}

\def\Ad{\mbox{\rm Ad}} \def\ad{\mbox{\rm ad}} 

\def\mathR{\R}

\def\a{\alpha }

\def\ew{\hspace*{-.5mm}}   \def\ppe{\hspace*{-1mm}}

\newcommand{\Fus}[6]{F_{{\scriptstyle #1}{\scriptstyle #2}}
  \hspace*{.3mm}\displaystyle{\Bigl[} \ew \begin{array}{ll} {\scriptstyle #3 }
  \ppe & {\scriptstyle #4} \ppe \\[-2mm] {\scriptstyle #5}\ppe &
  {\scriptstyle #6}\ew \end{array}\displaystyle{\Bigr]}}
\newcommand{\CG}[6]{\displaystyle{\Bigl[} \,\ew \begin{array}{lll} 
  {\scriptstyle #1} \ppe
  & {\scriptstyle #2} \ppe & {\scriptstyle #3} \ew \\[-2mm] {\scriptstyle
  #4} \ppe & {\scriptstyle #5}\ppe & {\scriptstyle #6} \ew\end{array}
  \displaystyle{\Bigr]}}

\def\Mat{{\mbox{\rm Mat}}}

\def\ik{{\sf k}}
\def\min{{\mbox{\rm min\/}}}
\def\cH{{\cal H}}
\def\cS{{\cal S}}
\def\itY{{\rm Y}}

\def\astk{{\, \stackrel{\ast}{,}\, }} 

\def\tf{{\sf f\, }} 
\def\tL{{\sf L}}  
\def\tA{{\sf A}}
\def\tB{{\sf B}}
\def\tF{{\sf F}}
\def\tY{{\sf Y}}
\def\tS{{\sf S}}
\def\tr{{\sf tr\ }}

\def\tCS{{\sf CS}}
\def\one{{\bf 1}}
\def\bigtimes{\hbox{$\,\diagup\hskip-11pt\diagdown\,$}}
      
\title{Brane Dynamics in Background Fluxes  \\  
        and Non-Commutative Geometry}
\author{ Anton Yu.\ Alekseev \\ Institute for Theoretical Physics, 
    Uppsala University \\ Box 803, S--75108 Uppsala, Sweden;
\email{alekseev@teorfys.uu.se} }
\author{ Andreas Recknagel \\ Max-Planck-Institut f\"ur 
              Gravitationsphysik, 
       Albert-Einstein-Institut \\ Am M\"uhlenberg 1, D--14424 Potsdam, 
         Germany;
\email{anderl@aei-potsdam.mpg.de}}
\author{ Volker Schomerus \\ II. Institut f\"ur Theoretische Physik, 
          Universit\"at Hamburg\\ Luruper Chaussee 149, 
           D--22761 Hamburg, Germany \\ and \\ 
         Institute for Advanced Study, School of Natural Sciences\\
         Einstein Drive, Princeton, New Jersey 08540;  
\email{vschomer@x4u.desy.de}}
\abstract{Branes in non-trivial backgrounds are expected to 
exhibit interesting dynamical properties. We use the boundary 
conformal field theory approach to study branes in a curved 
background with non-vanishing Neveu-Schwarz 3-form field strength.
For branes on an $S^3$, the low-energy effective action is
computed to leading order in the string tension. It turns 
out to be a field theory on a non-commutative `fuzzy 2-sphere' 
which consists of a Yang-Mills and a Chern-Simons term. We 
find a certain set of classical solutions that have no 
analogue for flat branes in Euclidean space. These solutions 
show, in particular, how a spherical brane can arise as bound 
state from a stack of D0-branes.}
\preprint{IASSNS-HEP-00/23 \\ AEI 2000-4 \\
hep-th/0003187}
\begin{document}
\section{Introduction}

Many important properties of D-branes can be understood in terms of 
open strings attached to the world-volume of the brane. 
At sufficiently low energies, the behavior of these open strings 
may be described through an effective field theory of the light 
modes from the open string spectrum. Typically, these field theories 
involve vector fields propagating on the brane along with a number 
of scalar fields that describe fluctuations of the brane in the 
transverse direction.  

\smallskip\noindent
It was discovered more recently that the presence of a background 
B-field causes the world-volume of the brane to become non-commutative. 
The initial observation for branes on a 2-torus \cite{DoHu,CDS} was clarified 
and generalized in a number of papers \cite{NCTor,ChHo,Vol,SeiWit}. One can 
see this non-commutative geometry of branes in their effective action. 
In fact, it naturally becomes a field theory on a non-commutative space so 
that the non-commutativity of the world-volume has direct influence on the 
dynamics of the brane.   

\noindent
Much of the existing work in this direction has been devoted to branes
in flat space.  A perturbative analysis along the lines of \cite{Vol}, 
on the other hand, shows that the quantization of world-volume geometries 
should be a much more general phenomenon which persists in the case of 
curved backgrounds. In fact, quantization of world-volumes appears to be  
inevitable in curved backgrounds because the string equation of motion 
enforce a non-vanishing NS 3-form field strength $H$ whenever the 
underlying space is not Ricci-flat. Since a non-zero field strength 
cannot possibly possess a vanishing 2-form potential $B$, branes in 
curved space are necessarily equipped with a non-zero B-field and hence 
they are quantized. 

\noindent The first rigorous analysis of brane non-commutativity in a 
curved space, namely on a 3-sphere, was presented in \cite{AlReSc}, 
using the world-sheet approach to D-branes. In that work, we gave 
a prescription to extract non-commutative brane world-volumes from 
the data of a boundary CFT. This definition builds on the closed string 
ideas of \cite{FG} (see also \cite{FGR} for examples) and is 
independent of notions from classical geometry, thus providing further 
evidence that brane non-commutativity is a general feature. 

\smallskip\noindent
An exact conformal field theory description is available for certain 
maximally symmetric branes on an $S^3$. It was shown in \cite{AlSc} that, 
classically, these branes are wrapped on conjugacy classes of 
${\rm SU(2)} \cong S^3$. The generic classes are 2-spheres in $S^3$, 
and due to the B-field they are quantized \cite{AlReSc} such that the 
algebra of `functions on the brane' is a full matrix algebra or some 
(mildly non-associative) q-deformation thereof, depending on the size 
of the 3-sphere in which the branes are embedded. The size of the 
matrix algebra is determined by the radius of branes themselves. 
These world-volume algebras are also known as `fuzzy-spheres', see 
\cite{Hop,Mad} and references therein. 

\noindent
As these fuzzy spheres are now embedded into a full string theory with 
all its dynamical degrees of freedom, we may wonder about the field 
theory that describes the light open string modes living on the 
fuzzy sphere. Field theories with the expected field content have
been described in the non-commutative geometry literature 
\cite{Mad,FFT,Wat,Madbook,Klim}. In particular, Klim\v c\'\i k 
\cite{Klim} has suggested a family of actions which are arbitrary 
linear combinations
of non-commutative Yang-Mills and Chern-Simons terms. The string theory 
effective action is of this type but selects a particular member of 
this family for which the mass terms present in the non-commutative 
Yang-Mills and Chern-Simons actions cancel each other. 

\smallskip\noindent
We will see that our field theories on the fuzzy sphere possess 
interesting classical solutions some of which have no analogue 
in effective field theories on flat branes. They 
describe stacks of spherical branes that evolve into superpositions
of spherical branes with different radii. In particular, a stack of 
$N$ D0-branes (i.e.\ $N$ spheres of vanishing radius) can condense into 
a single D2-brane of radius $r \sim N$. This dynamical generation of 
new world-volume dimensions occurs at the expense of a partial breaking 
the gauge symmetry.

\noindent The formation of extended objects from smaller constituents 
has of course been discussed extensively in the context of M-theory (see 
e.g.\ \cite{Tay} and references therein) and also for branes in constant 
RR-background fluxes \cite{Mye}. While the latter work takes the 
Born-Infeld action as a starting point, our analysis is based on 
an exact open string background given by a boundary CFT. 
In this way, we are also able to show that the effects in question 
are inherited from the brane's non-commutativity.

\noindent 
This requires, however, to understand the brane dynamics 
beyond small fluctuation. The latter were studied recently in 
\cite{BaDoSc,Paw} in order to establish the stability of branes on 
group manifolds. Many of the statements we will make below can 
easily be extended to other group manifolds, too. We restrict 
ourselves to the group SU(2) for simplicity and because the background 
${\rm SU(2)} \cong S^3$ is also an important ingredient of the CFT 
formulation of the Neveu-Schwarz 5-brane, see e.g.\ \cite{fivebr}, 
such that our findings may be useful to study the geometry of D-branes 
in the presence of a stack of 5-branes. Similarly, our SU(2) WZW 
results are applicable in the study of branes on an AdS$_3 \times 
S^3$ string background, see e.g.\ \cite{GKS}.

\medskip\noindent
The paper is organized as follows: In the next section we 
will briefly list a few facts about branes, B-fields and non-%
commutative geometry of branes in flat space, including a few 
important formulas that are rather convenient to prepare the ground 
for the following less standard discussion. Section 3 is mainly a review 
of the results in \cite{AlSc,AlReSc}. It contains all the results
we shall need for our computation of the effective action, which is 
carried out in Section 4. For simplicity, this will be illustrated 
in the bosonic theory first, but we will sketch how to extend the 
calculation to full supersymmetric string backgrounds like a NS 5-brane. 
It will turn out that the (bosonic part of the) low-energy effective 
action is the same as for bosonic strings on $S^3$. 
Section 5 contains a detailed analysis of classical solutions
and their interpretation. Finally, we end with conclusions and 
some open problems.

\section{Flat branes and non-commutative gauge theory} 

Before we show how to calculate low-energy effective actions for 
branes wrapping an $S^2 \subset S^3$, let us briefly review a few 
important steps of the analogous calculation in the standard 
case of branes in flat $n$-dimensional Euclidean space $\mathR^n$, 
or on a flat torus $\T^n$. Consider a D-brane which is localized 
along a $p$-dimensional hyper-plane $V_p$ in the target, with 
tangent space $TV_p$. The conformal field theory associated with 
such a Euclidean D-brane is defined on the upper half of the 
complex plane. It contains an $n$-component free bosonic field 
$X= (X^\mu(z,\bar z)),\ \mu = 1,\dots,n,$ subject to Neumann 
boundary conditions in the directions along $TV_p$ and Dirichlet 
boundary conditions for components perpendicular to the world-volume 
of the brane. From the free bosons, one may obtain various new fields, 
in particular the vertex operators for the open string tachyons  
$$ 
V_k (x) = \ :\!\exp(i k X(x))\!:  \ \quad \mbox{ for all } \ \ \ 
    k \in TV_p \ \ ,
$$
which can be inserted at any point $x$ on the real line. These operators 
correspond to the lightest states $|k\rangle$ within each sector of 
fixed momentum $k$. One can think of them as being assigned to the 
eigenfunctions $e_k = \exp (i k u)$ of the momentum operator on the 
world-volume $V_p$ of the brane. To make this more explicit we will 
use the notation $V[e_k](x) = V_k(x)$.     
  
\noindent
When there is a magnetic field $B = (B_{\mu \nu})$ on the brane, the 
operator product expansion (OPE) of these U(1)-primaries reads  
$$ 
V[e_{k_1}](x_1) \, V[e_{k_2}](x_2) \ = \ \frac{1}{(x_1 - x_2)^{2\a' 
    k_1 \cdot k_2}} e^{- i \frac{\pi \a'}{2}\; k_1^{\rm t} \Theta  k_2} 
    \;V[e_{k_1+k_2}](x_2) 
   + \ldots\ \ , 
$$ 
up to contributions involving descendant fields. In writing such  
OPEs of boundary operators we assume throughout the whole paper that 
their insertion points are ordered $x_1 > x_2$. 
The product $k_1 \cdot k_2 = G^{\mu \nu} k_{1,\mu} k_{2,\nu}$ is to be 
taken with the inverse of the open string metric $G$ rather than of the 
original closed string metric $g_{\mu \nu}$. $G^{\mu\nu}$ can 
be read off from the standard expression $h =\ap G^{\mu \nu} k_\mu 
k_\nu= 1/2 \, G^{\mu \nu} (\sqrt{2\ap} k_\nu) (\sqrt{2\ap} k_\mu) $ for 
the scaling dimension of the boundary primary fields $V[e_k]$. The 
relation of $\Theta$ and $G$ to the closed string parameters $B$ and 
$g = 1$ is given by \cite{Vol}
\be
\Theta \ = \ \frac{2 }{B-B^{-1}}\ \ \ \mbox{ and } \ \ \ 
G \ = \  \frac{1}{1-B^2}\ \ .  
\label{magic}
\ee
In a particular limit (`zero slope limit' \cite{SeiWit}) of the theory for
which $\a' G^{\mu \nu}$ becomes zero, the boundary fields have vanishing 
conformal dimension and the operator product expansions are no longer 
singular. Thus we obtain 
$$ 
V[e_{k_1}](x_1) \, V[e_{k_2}](x_2) \ = \ 
      V[e_{k_1} \ast e_{k_2}](x_2)\ \ .  
$$ 
The product $e_{k_1} \ast e_{k_2} = \exp (- i \frac{\pi \a'}{2}\; k_1^{\rm t} 
\Theta  k_2) \, e_{k_1} e_{k_2}$ is the Moyal-Weyl product of functions 
on $V_p$ (see \cite{Vol,SeiWit} for details). We can pass to 
a more universal form that does not make reference to the basis
$e_k$ in the space of functions: If we introduce $V[f] = 
\sum \hat f_k\,  V[ e_k]$ with $\hat f_k$ being the Fourier 
coefficients of $f$, we obtain 
$$ 
V[f] (x_1) \ V[g] (x_2) \ = \ V [ f \ast g ] (x_2)\ \ .
$$
This formula holds for arbitrary functions $f,g$ as long as we 
are in the zero slope limit. 
In order to compute the low-energy effective action, we need two 
further operator products involving the boundary current $j^\mu(x)$, 
namely          
\ba 
j^\mu (x_1) \ j^\nu(x_2) \ & =  & \frac{\ap}{2}\,
    \frac{ G^{\mu\nu}}{(x_1-x_2)^2}
   \ + \ \dots \ \ , \label{ope1}\\[2mm]
j^\mu (x_1) \, \ V[f] (x_2) & = & \frac{\ap \, 
  G^{\mu\nu}}{x_1-x_2}\ \ V[-i \partial_\nu f] (x_2) \ + \ 
\dots\ \ . \label{ope2} 
\ea
With the help of these equations, the $n^{th}$ order terms 
for massless fields in the LEEA are obtained from an $n$-point
function of operators $:\!\!j^\mu V[A_\mu]\!\!:\!(x)$ where $A_\mu$ 
is the (possibly non-abelian) gauge field, i.e.\ $A_\mu(u)$ are 
functions on the world-volume of the brane, which take values 
in $\Mat(N)$ for a stack of $N$ identical branes. 
In the zero slope limit, the resulting action is given by 
a gauge theory on the non-commutative plane or torus \cite{DoHu,SeiWit}, 
$$ \cS(A)_{\rm flat} \ = \ \frac{1}{4} \, \int  d^p u  
   \  F_{\mu \nu} \ast F^{\mu \nu}
$$ 
where $F_{\mu \nu}(A) = \partial_\mu A_\nu - \partial_\nu A_\mu 
+ i [A_\mu \ast A_\nu]$ and where $\ast$ is the Moyal-Weyl product; 
the integration extends over the world-volume of the brane.

\section{Branes on $\mathbf{S^3}$ and fuzzy spheres} 

Strings moving on a three-sphere are described by the SU(2) WZW model at 
level $\ik$. Here, the radius $R$ of $S^3$ is related 
to the level by  $R \sim \sqrt{\a' \ik}$. Because $S^3$ is curved, the 
string equations of motion enforce the presence of a constant NS 
3-form field strength $H  \sim \Omega/\sqrt{\a' \ik}\;\,$ where 
$\Omega$ is the volume form on the unit sphere. 

\noindent 
The world-sheet swept out by an open string in $S^3$ is 
parametrized by a map $g: {\rm H} \rightarrow {\rm SU(2)}$ from the 
upper half-plane H into the group manifold SU(2)$\, \cong S^3$. From 
this field $g$ one obtains Lie algebra valued chiral currents 
$$ J(z) \ = \ - \ik \, (\partial g) g^{-1} \ \ \ , \ \ \ \
   \bar J(\bar z) \ = \ \ik \, g^{-1} \bar\partial g \ \  $$
as usual. There exists an exact conformal field theory description for
maximally symmetric D-branes on SU(2), which are characterized by 
the gluing condition $J(z ) = \bar J (\bar z)$ along the boundary 
$z = \bar z$. The geometry of the associated branes is encoded in 
these gluing conditions \cite{AlSc}. First of all, they force the 
endpoints of open strings to stay on conjugacy classes of SU(2). 
The generic such classes are 2-spheres, except for two cases 
where the classes degenerate into the single points $e$ resp.\ 
$-e$, where $e$ is the unit element of SU(2). 
Moreover, the condition $J = \bar J$ selects a particular two form 
potential $B$ (`B-field') on the brane, which is given by  
$$
B \ = \ \frac{\Ad(g)+1}{ \Ad(g)-1} \ .
\label{Bfield} 
$$
Here $Ad(g)$ denotes the adjoint action of the SU(2) on the tangent 
space of the brane. The existence of a non-vanishing B-field causes
the brane world-volume to become non-commutative. From the previous 
analysis of branes in flat space we know that it is the antisymmetric 
tensor $\Theta$ rather than the B-field itself (or its inverse) 
that describes the semi-classical limit of the brane's non-commutative 
world-volume algebra. $\Theta$ is computed from $B$ by means
of the formula (\ref{magic}), leading to  
\be 
\Theta \ = \ \frac{2}{B-B^{-1}} \ = \ \frac{1}{2} 
     \left(Ad(g) - Ad(g)^{-1}\right) \ \ . 
\label{Theta} 
\ee  
The semi-classical extension of the above analysis shows that, for 
fixed gluing conditions, only a finite number of SU(2) conjugacy 
classes satisfy a Dirac-type flux quantization condition \cite{AlSc,Gaw}. 
These `integer' conjugacy classes are the two points $e$ and $-e$ 
along with $\ik - 1$ of the spherical conjugacy classes, namely those 
passing through the points $\;{\rm diag}(\exp(2\pi i \alpha/\ik),\exp(-2\pi i 
\alpha/\ik))$ for $\alpha =\frac{1}{2},\ldots,\frac{\ik-1}{2}$.

\medskip\noindent
Our discussion of brane dynamics in the next section will be 
based upon a computation of the leading non-vanishing terms of the 
low-energy effective action around the point $1/\a'k = 0$. For this 
reason, let us briefly look at the classical geometry in the limit 
$\a'\ik \to \infty$ in which the three-sphere grows and approaches 
flat 3-space. If one parametrizes $g$ by $X$ taking 
values in the Lie algebra su(2), such that $g \approx 1 - X$, the 
expression for $\Theta$ becomes $\Theta = \ad(X)$. 
This is the standard Kirillov bi-vector on the spheres in the
algebra su(2)$\, =\, \R^3$. Consequently, the geometry of the 
limiting theory $\a'\ik = \infty$ is very close to the well-known 
situation of flat branes in a flat background with constant 
B-field, and we expect that the world-volume algebras of our 
branes in the WZW model will be quantizations of two-spheres
equipped with their standard Poisson structure. According to 
Kirillov's theory, all `integer' 2-spheres are quantizable 
and in one-to-one  correspondence with irreducible 
representations of SU(2). Kirillov's theory implies that `functions' 
on the quantized 2-spheres form matrix algebras $\Mat(2\a+1)$ 
of size $2 \alpha + 1$ where $\alpha$ can be any of the numbers 
$\a = 0,1/2,1, \dots$ In this way we have identified our branes 
with representations of SU(2) such that the brane's label
$\alpha$ - which is proportional to its radius -- denotes the spin 
of the associated representation. 
   
\bigskip\noindent
In addition to the geometric picture developed so far, we will now 
list a number of exact results from the boundary conformal field 
theory description \cite{Car} of the associated quantum theory of 
open strings. Their derivation is explained in \cite{AlReSc}. The 
chiral fields of the SU(2) WZW model form an affine Kac-Moody algebra 
denoted by $\widehat{\rm SU}(2)_\ik$. It is generated by fields 
$J_a(z), a = 1,2,3,$ which possess the following operator product 
expansions
\be  
J_a (x_1) \ J_b(x_2)\  = \ \frac{\ik}{2}\, \frac{ \delta_{a b}}{(x_1-x_2)^2}
                   \ + \  \frac{i\, f_{ab}^{\ \ c}}{(x_1-x_2)} \, J_c(y) 
                        \ + \ \dots \ \  \ \label{JOPE1} 
\ee
where $f$ is the Levi-Civita tensor. 
The open string vertex operators of the D-brane theories we are
dealing with are organized in representations of this symmetry 
algebra. For the brane with label $\alpha =0,\frac{1}{2},
\ldots,\frac{\ik}{2}$, there appear only 
finitely many such representations with integer angular momentum 
between $J = 0$ and $J = \min(2\a, \ik- 2\a')$. Each 
representation $J$ contains ground states of lowest possible 
energy, which form $(2J+1)$-dimensional multiplets spanned 
by some basis vectors $Y^J_m$ with $|m|< J$. To these states we 
assign vertex operators $V[Y^J_m](x)$. They are similar to the tachyon 
vertex operators $V[e_k]$ for open strings in flat space. While the 
latter are associated with eigenfunctions of linear momentum $k$, 
our vertex operators $V[Y^J_m]$ carry definite angular momentum,  
and one may think of the elements $Y^J_m$ as spherical harmonics on 
the 2-sphere that forms the classical world-volume of the brane $\a$. 
Because of the quantization, 
there appears a cutoff on the angular momentum, i.e.\ $J$ is 
restricted by $J \leq \min(2\a, \ik- 2\a')$. The operator 
product expansion of these open string operators was found
to be of the form \cite{Run,AlReSc}
\be \label{boundOPE}
    V[Y^I_i](x_1)\ V[Y^J_j](x_2) \ = \ {\sum}_{K,k}\ x_{12}^{h_K - h_I -
    h_J} \ \CG{I}{J}{K}{i}{j}{k}\ \Fus{\alpha}{K}{J}{I}{\alpha}
    {\alpha} \ V[Y^K_k](x_2)\ + \dots\     
\ee
where $h_J$ is the conformal dimension of $V[Y^J]$, $[:::]$ denote 
the Clebsch-Gordan coefficients of the group SU(2), and $F$ stands for 
the fusing matrix of the WZW-model. In the limit 
$\ik \rightarrow \infty$, the fusing matrix elements approach
the $6J$ symbols of the classical Lie algebra ${\rm su}(2)$. At the
same time the conformal dimensions $h_J = J(J+1)/(\ik+2)$ 
tend to zero so that the OPE (\ref{boundOPE}) of boundary fields 
becomes regular as in a topological model. 
\newline\noindent 
For the limiting theory there exists a much more elegant way of 
writing the operator product expansion 
(\ref{boundOPE}). To see this observe that SU(2) acts on the space 
of $(2\a+1) \times (2\a +1)$-matrices by conjugation with group 
elements evaluated in the $(2\a+1)$-dimensional representation. 
Under this action, the $(2\a+1)^2$-dimensional space $\Mat(2\a+1)$ 
decomposes into a finite set 
of SU(2) modules of spin $J = 0 ,1 , \dots, 2\a$. This is precisely the 
list of representations which appears in the open string theory of our 
brane $\a$ when the level is sent to infinity. Hence, we can 
identify the ground states $|Y^J_m\rangle $ with matrices 
$\itY^J_m \in  \Mat(2\a+1)$. It is a remarkable feature of this 
identification that the multiplication of the matrices $\itY^I_i$ and 
$\itY^J_j$ turns out to encode the information about the operator products 
in our conformal field theory, 
$$     
 \label{boundOPE1}
    V[\itY^I_i](x_1)\ V[\itY^J_j](x_2) \ = \ V[\itY^I_i \ast \itY^J_j](x_2)
    \ \ \ .  
$$
The emergence of the matrix product $\ast$ in this relation was  
expected from our geometric analysis above. It shows that the 
world-volumes of branes in $S^3$ become fuzzy two-spheres in the 
quantum theory. 
We can now rewrite the operator product in a way which does no longer 
refer to a particular choice of a basis. In fact, for an arbitrary matrix 
$\tA \in \Mat(2\a+1)$ with $\tA = \sum a_{J m} \itY^J_m$ we introduce
$V[\tA] = \sum a_{Jm} V[\itY^J_m]$ and obtain 
\be \label{boundOPE2}
    V[\tA_1](x_1)\ V[\tA_2](x_2) \ = \ V[\tA_1 \ast \tA_2](x_2)\ \ \   
\ee
for all $\tA_1,\tA_2 \in \Mat(2\a+1)$. This product allows us to compute 
arbitrary correlations functions of such vertex operators, 
\be \langle V[\tA_1](x_1) \ V[\tA_2](x_2) \ \cdots \ 
  V[\tA_n](x_n) \rangle \ = \ \tr (\tA_1 \ast \tA_2 \ast \cdots \ast \tA_n)\ \ .
\label{wzwcorr} \ee
The trace appears because the vacuum expectation value is  
SU(2) invariant and the trace maps matrices to their SU(2) invariant 
component.  

\smallskip\noindent
Let us close this section on the conformal field theory of branes 
on $S^3$ by displaying a further OPE, namely the one that involves a 
current $J^a$ along with the vertex operator
$V[\itY^J_j]$, 
\be\label{Ldef}
J_a (x_1) \ V[\itY^J_j](x_2) = 
\frac{1}{x_1-x_2} \ V[ L_a \itY^J_j] (x_2) \ + \ \dots \ 
\mbox{ with } \ 
   \ L_a \tA \ := \ [\, \itY^{1}_a \astk \tA\, ] 
\ee
for all matrices $\tA \in \Mat(2\a+1)$. The operator $L_a$ expresses 
the action of the angular momentum on spherical harmonics.

\section{Low-energy effective field theory} 

With the previous description of the world-sheet theory as a firm 
basis, we want to address the main aim of the paper, namely the 
construction of the effective action that describes D-branes wrapped 
on an $S^2 \subset S^3$ at low energies. The three-spheres, or 
rather the SU(2) WZW models, occur as internal spaces in (super) 
string backgrounds like AdS$_3 \times S^3 \times T^4$ or in the 
world-sheet theory for the background with $\ik-2$ Neveu-Schwarz 
5-branes, which provide a source for $\ik-2$ units of NS 3-form flux 
through a three-sphere surrounding their (5+1)-dimensional world-volume. 
The CFT description of such a NS 5-brane background involves a tensor 
product of 5+1 free bosons, of a linear dilaton and of an SU(2) WZW 
model \cite{fivebr}. For our purposes, it will be sufficient to 
approximate the `external' part $M^7$ of these targets by a flat 
space-time. At the moment, we furthermore restrict ourselves to 
bosonic strings, but it will turn out that the results for the 
low-energy effective action carry over the supersymmetric case. 

\noindent Thus we consider stacks of $N$ D$p$-branes which span 
a certain flat world-volume in $\R^7$ and  wrap a (fuzzy) two-sphere 
with label $\alpha$ in the $S^3$-component of space-time. The 
massless field of this brane configuration fall into two groups 
which we want to discuss separately. 

\smallskip\noindent
First there are the vector bosons and scalars which are associated 
with the seven external directions transverse to the $S^3$. When we 
compute their contribution to the action, we have to use vertex 
operators $:\! j^\mu V[A_\mu] V[\tA]\!\!:\!(x)$. Here, $\mu = 0, \dots, 6$
labels the coordinates transverse to the $S^3$, and $A_\mu$ is either 
a U($N$) gauge or a scalar field on the brane depending on whether the 
direction of $\mu$ is parallel or perpendicular to the brane. The 
third factor $V[\tA]$ in the vertex operator corresponds to the 
ground states for the brane on $S^3$, which means that $\tA$ is a 
matrix of size $2\a+1$. The actual computation of the associated 
terms in the effective action proceeds exactly as for branes in flat 
space, except for the contribution from ground states
in the WZW factor of the theory. But the evaluation of the latter  
is completely independent of the first seven directions, and our 
formula (\ref{wzwcorr}) shows that the WZW model contributes precisely 
in the same way as usual Chan-Paton factors do. Hence, the effective
action for the fields $A_\mu$ appears as an ordinary non-abelian 
Yang-Mills-Higgs theory on a D$(p-2)$-brane. The effect of the 
two dimensions which wrap the $S^2 \subset S^3$ is to replace   
U($N$) by the larger gauge group U($N\,(2\a +1)$). In particular, 
non-abelian gauge theories on the world-volume of a single brane 
appear for $\alpha > 0$.  

\medskip\noindent
The discussion of the remaining three fields that come with the $S^3$ 
(scalars viewed from the external part of the brane world-volume)  
requires a lot more effort. To begin with, let us rescale the currents 
of the WZW model so as to bring their operator product expansions in 
a form similar to eqs.\ (\ref{ope1},\ref{ope2}). As in the flat case, 
we extract the inverse of the open string metric from the conformal 
dimensions, here given by $h_J= J(J+1)/(\ik+2)$. To leading order 
in $1/\ik$, this yields the metric 
\be    G^{a b} \ = \ \gamma\,  \delta^{a b} \ \
   \ \mbox{with } \ \ \ \gamma = 2/\ik \ \ .
\ee
For the Schwinger term in the OPE of WZW currents $J_a$ to resemble 
eq.\ (\ref{ope1}), we introduce $j^a = (\sqrt{2 \ap}/\ik) J_a$. The 
operator product expansions of these rescaled currents are 
\ba 
j^a (x_1) \; j^b(x_2)\ & = & \frac{\ap}{2}\, \frac{ G^{a b}}{(x_1-x_2)^2}
                   \ + \ \ap\, \frac{ i \, \tf^{a b}_{\ \ 
                     c}}{(x_1-x_2)} \,  j^c(y) 
                        \ + \ \dots \ \ \ \ \label{jOPE1} \\[2mm]
j^a (x_1) \; V[\tA] (x_2) & = & \frac{\ap \, G^{ab}}
     {x_1-x_2}  \ V[\tL_b \tA] (x_2) \ + \ \dots  \label{jOPE2}
\ea 
where we also use a rescaled angular momentum operator 
$\tL_a = L_a /\sqrt{2\ap}$. The structure constants $\tf$ in the first 
line are obtained from the structure constants $f$ of the classical 
Lie algebra ${\rm su(2)}$ through $\tf_{\!a b}^{\ \ c} = f_{a b}^{\ \ c}/
\sqrt{2\ap}$. Here and in the following, indices $a,b,c \dots$ on 
the objects $\tL$ and $\tf$ are raised and lowered with the open 
string metric $G$. In particular, this implies that $\tf^{a b}_{\ \ c} 
= 2 f^{a b }_{\ \ c}/ (\sqrt{2\ap} \ik)$.      

\smallskip\noindent
Let us focus on the WZW boundary fields $:\!j^a V[\tA_a]\!:(x)$ 
where $\tA_a = \sum a^{Jm}_a \itY^J_m$ is an arbitrary set of matrices 
subject to the physical state condition $\tL^c \tA_c = 0$; summation 
over the three directions $a = 1,2,3$ tangent to the 3-sphere is understood. 
For simplicity, we shall omit the simple factors $V[f]$ that come with the 
seven transverse directions throughout most of our discussion. They are 
taken into account if we consider $\tA_a$ as matrix valued functions on 
the world-volume of the brane rather than constant matrices. $\tA_a$ may 
be regarded as three massless bosons that live on our fuzzy sphere (or the 
$(p+1)$-dimensional world-volume of the brane). In the limit $\ik \to \infty$ 
they can also be viewed as massless fields depending on the $p-1$ brane 
coordinates which are transverse to the $S^3$ because the conformal 
dimensions $h_{I}$ of the ground states $|\itY^I_i\rangle$ vanish as we 
send the level to infinity. When we want to describe a stack of $N$ of 
our $\alpha$-branes, the coefficients $a^{Jm}_a$ of the fields $\tA_a$ 
have to be $N \times N$ matrices and hence $\tA_a \in \Mat(N) \otimes 
\Mat(2\a +1)$. We stress, however, that the operators $\tL_a$ act 
exclusively on the second tensor factor and not on the Chan-Paton degrees 
of freedom.     

\noindent
We are now prepared to compute the effective action for the three fields 
$\tA_a$ to leading order in $\a'$. As in the case of flat branes it 
turns out that only 3- and 4-point functions contribute to this order of 
the open string scattering amplitudes. For the 3-point function one 
finds the following expression 
\ba   
 & & \hspace*{-0.5cm} \langle\; :\!j^a V[\tA_a]\!: (x_1) \, :\!j^b V[\tA_b]\!: 
     (x_2)\, :\!j^c V[\tA_c]\!: (x_3)\; \rangle\label{3pt} \\[2mm] 
 & = &   \frac12\, \frac{(\ap)^2}{x_{23}x_{12} x_{31}}
\left[ \ \frac{x_{23}}{x_{12}}\,  G^{ab} \, \tL^{c,(1)}
        - \frac{x_{31}}{x_{12}}\, G^{ab}\, \tL^{c,(2)}
         +\frac{x_{31}}{x_{23}} \, G^{bc}\, \tL^{a,(2)}
         -\frac{x_{12}}{x_{23}} \, G^{bc}\,  \tL^{a,(3)}
             \right.\nn \\[2mm]
   & &  \left.   + \frac{x_{12}}{x_{31}}\, G^{ac}\, \tL^{b,(3)}
        - \frac{x_{23}}{x_{31}} \, G^{ac}\, \tL^{b,(1)}
             - i \,\tf^{abc} \ \right]
   \langle V[\tA_a] \ V[\tA_b] \ V[\tA_c]  \rangle
\ + \ O((\ap)^3)\nn\ \ . 
\ea
Here, the superscript $\,{}^{(j)}$ means that the operator acts on 
the argument of the $j^{th}$ vertex operator in the correlation 
function. The terms involving $\tL$ arise from contracting two 
currents into a Schwinger term and acting on one of the primary fields 
with the third current. 
The structure constants $\tf^{abc}$ appear upon contraction of two 
currents into the second term of eq.\ (\ref{jOPE1}) and subsequent 
contraction of the resulting current with the third current in the 
correlator. In the bosonic situation, there are higher order terms 
which we did not spell out because they will not be needed below.

\noindent
{}From formula (\ref{3pt}), we recover the corresponding expression 
for the flat space theory by the replacements 
$\tL^a \rightarrow - i \partial^a$ and, more importantly, $\tf^{abc} 
\rightarrow 0$. The 3-point functions feel the non-abelian nature 
of the theory through the terms that involve the Lie algebra structure 
constants. In the 4-point functions, such terms are suppressed by at 
least one order of $\ap$, and the leading $O((\ap)^2)$ contributions 
come from pairwise contractions of two currents into a Schwinger term. 
They are identical to the corresponding terms in the case of flat 
D-branes except for the replacement $\tL^a \rightarrow -i\partial^a$. 

\noindent
Before we can combine this information into some effective
action, we need to consider the mass terms of the theory. Recall
that the mass of open string modes is determined from the
conformal dimensions in the internal sector of the theory
through $M^2 = (h-1)/\ap$. The modes we consider have conformal 
dimension $h = h_J + 1$ so that, to leading order in $1/\ik$, 
\be
\label{massformula}
M^2 = \frac{1}{\ap}\, h_J \ = \  \frac{J(J+1)} {\ap\ik}\ \ .
\ee  
This relation shows that it is the combined limit $\ap \to 0$ 
and $\ap\ik\to\infty$ that is relevant for our low-energy 
considerations. The values in (\ref{massformula}) constitute the 
spectrum of the operator $\frac{2}{\ik}\sum_a \tL_a \tL_a  =  
\tL_a \tL^a$ so that the formula for $M^2$ results in the following 
mass term for the Lagrangian of our theory
$$       
\cS_0 \ = \  \frac12\ \tr \left(\, \tA_a \ast \tL_b \tL^b
               \tA^a \, \right)
      \ = \ - \, \frac12\, \tr \left(\, \tL_b \tA_a \ast \tL^b
               \tA^a \, \right) \ \ .
$$
This is the free theory to which we add the interaction terms encoded 
in the 3- and 4-point functions. They are obtained as usual, with 
the help of the physical state condition $\tL^a \tA_a = 0$, by 
multiplication with ghost propagators and after summation over all 
permutations of the insertion points $x_i$. The result is 
\vspace*{2mm} 
\ba
\cS^{(3)} & = & - \, \frac12 \ \tr \left(\, \tL_a \tA_b \ast
                [ \tA^a \astk \tA^b ]    
              + [ \tA_a \astk \tA_b ] \ast \tL^a \tA^b
              - \frac{2 i}{3} \, \tf^{a b c} \tA_a \ast
                 [ \tA_b \astk \tA_c ]\,\right)\ \ ,  \nn \\[2mm]
\cS^{(4)} & = & -\, \frac{1}{4} \ \tr \left(\,[ \tA_a \astk
               \tA_b ] \ast [ \tA^a \astk \tA^b ]\,\right)\ \ .
                \nn \vspace*{4mm} 
\ea
Up to their normalization, the formulas for $\cS^{(3)}$ and $\cS^{(4)}$
are obvious from eq.\ (\ref{3pt}) and our description of the leading
contribution to the 4-point function. The relative normalization of 
$\cS^{(3)},\cS^{(4)}$ and $\cS_0$ may be determined most easily by 
comparison with the abelian case $\tf_{a b c} = 0$.   
The full effective action 
\be
\label{act}
 \cS =  \cS_0 + \cS^{(3)} + \cS^{(4)} 
\ee
for the scalar fields on our brane in $\R^7 \times S^3$ can be 
brought into a 'geometric' form by introducing a field strength  
\be \label{fieldstr}
  \tF_{a b}(\tA)\, = \,   
   i\, \tL_a \tA_b - i\, \tL_b \tA_a + i \,[ \tA_a \astk \tA_b] 
   +  \tf_{a b c } \tA^c
\ee
and a non-commutative analogue of the Chern-Simons form 
\be  \label{CSform}
  \tCS_{a b c}(\tA)\, = \, \tL_a \tA_b \ast \tA_c 
                   + \frac{1}{3}\; \tA_a \ast [ \tA_b \astk \tA_c]
                   - \frac{i}{2}\; \tf_{a b d}\; \tA^d \ast \tA_c \ \ .
\ee
Then our effective action can be written as sum of a Yang-Mills theory  
and of a Chern-Simons theory on the fuzzy sphere, 
\be  \label{geomeffact}
  \cS\ =\  \cS_{{\rm YM}} + \cS_{{\rm CS}}\ = \
      \frac{1}{4}\ \tr \left( \tF_{a b} \ast \tF^{a b} \right) 
             - \frac{i}{2}\  \tr \left( \tf^{a b c}\; \tCS_{a b c} \right)
\ee  
In order to render the effective action dimensionless, the expression
on the right hand side is to be multiplied with the appropriate 
power $(2\pi\ap)^2$ of the string tension. 

\noindent In the computation leading from (\ref{act}) to (\ref{geomeffact}) 
we have added a term of second 
order in the fields $\tA_a$ that vanishes on-shell, i.e.\ upon using the 
physical state condition $\tL^a \tA_a = 0$. The main motivation
for this modification is that it renders the action off-shell 
gauge invariant. In fact, some cumbersome but straightforward 
manipulations show that 
$$ \delta_\Lambda \, \cS_{\rm YM}(\tA) \ = \
   \delta_\Lambda \,\cS_{\rm CS}(\tA) \ = \ 0 \ \ \ 
   \mbox{ where } \ \ \ 
            \delta_\Lambda \, \tA_a \ = \ i\, \tL_a \Lambda \, +\,
            i\, [\, \tA_a, \Lambda\, ]    
$$
for arbitrary $\Lambda \in \Mat(N) \otimes \Mat(2\a+1)$. Note, in 
particular, that the 'mass term' in the Chern-Simons form (\ref{CSform}) 
is required by gauge invariance. On the other hand, the effective action 
(\ref{geomeffact}) is the unique combination of $\cS_{\rm YM}$ and 
$\cS_{\rm CS}$ in which mass terms cancel. As we shall see below, 
it is this special feature of our action that allows solutions 
describing translations of the branes on the group manifold. The 
action $\cS_{\rm YM}$ 
was already considered in the non-commutative geometry literature 
\cite{Madbook,Wat}, where it was derived from a Connes spectral triple 
and viewed as describing Maxwell theory on the fuzzy sphere. 
Arbitrary linear combinations of non-commutative Yang-Mills and 
Chern-Simons terms were considered in \cite{Klim}.  

\smallskip\noindent
There exists a nice way to rewrite our actions such that the statements 
about gauge invariance become completely obvious. To this end we note 
that the covariant derivative $\tL_a + [\tA_a,.\,]$ can be implemented as 
a commutator with the element 
\be\label{bdef} \tB_a \ = \ \tY_a + \tA_a  \ \ \ \ \mbox{ where } \ \ \ 
   \tY_a \ = \ \frac{1}{\sqrt{2\a'}} \, \itY^1_a \ \ ,  \ee
see eq.\ (\ref{Ldef}). The behavior of $\tB_a$ under gauge transformations 
is given by $\delta_\Lambda B_a = i [B_a, \Lambda]$. This property guarantees
the gauge invariance of $\cS_{\rm CS}$ and $\cS_{\rm YM}$ once we observe 
that both actions can be expressed entirely through $\tB_a$ according to  
\ba 
\cS_{\rm YM} & = & \frac{1}{4} \, \tr \left( \tF_{a b} \ast \tF^{a b} \right) 
\ \ \ \ \ \mbox{with} \ \ \  
\tF_{ab}  \ = \  i\,[\, \tB_a \astk \tB_b\, ] + \tf_{a b c}\, \tB^c \\[2mm]   
\cS_{\rm CS} & = & -\frac{i}{2}\,  \tr \left(\, 
    \frac13\,\tf^{a b c}\,  \tB_a \ast \tB_b \ast \tB_c  
    - \frac{i}{\a'\ik}\  \tB_a \ast \tB^a    
          + \frac{i}{3\a'\ik}\  \tY_a \ast \tY^a   \, \right)\ \ . 
\ea 
The two formulas are rather easy to verify by inserting the definition 
(\ref{bdef}) of $\tB_a$ and using the algebraic properties of $\tY_a$. 
The full low-energy effective action for branes in $S^3$ then takes 
the simple form 
\be
\label{niceeffact}
\cS = \tr \left( 
    - \frac14 \,[\, \tB_a \astk \tB_b\, ]\ast [\, \tB^a \astk \tB^b\, ]
    + \frac{i}3\, \tf^{a b c}\,\tB_a  \ast[\, \tB_b \astk \tB_c\, ]
    + \frac1{6\ap\ik} \, \tY_a \ast \tY^a \, \right)\ \ .
\ee

\bigskip\noindent
Up to now, we have only discussed SU(2) branes in bosonic string 
theory, which are unstable due to the presence of tachyonic modes. 
In the remainder of this section, we provide the formulas necessary 
to extend the bosonic discussion from above to the superstring case. 
As before, we will mostly ignore the curvature of the external 
manifold $M^7$ and pretend that our target has the form $\R^7 \times 
S^3$. In the supersymmetric situation, the SU(2) WZW model is supplemented 
by three free fermions $\psi^a(x)$, $a=1,2,3$, with the following 
operator product expansions (we use CFT conventions as in eq.\ 
(\ref{JOPE1}), string units are obtained by rescaling with $\ap$ 
in the same way as before) 
$$
\psi^a(x_1) \psi^b(x_2)  =  \frac{\ik}{2}\,\frac{\delta^{ab}}{x_1-x_2} 
+ \dots \ \ ,\quad\ 
J^a(x_1) \psi^b(x_2) =  \frac{i f^{ab}_{\ \ c}}{x_1-x_2}\; \psi^c(x_2) \ \ .
$$
The world-sheet supercurrent is given by (see e.g.\ \cite{Fuchs}) 
$$
G = G^7 + G_{\rm SWZW} = G^7 + \frac{2}{\ik}\; \delta_{ab}\, J^a \psi^b 
  + \frac{4i}{3\ik^2}\;f_{a b c}\, \psi^a \psi^b \psi^c \ \ .
$$
$G^7$ depends on the specific superstring background under consideration. 
E.g.\ for the NS 5-brane, we have \cite{fivebr} 
$$ 
G^7 = \eta_{\mu\nu}\,J^\mu \psi^\nu + J^6 \psi^6 
                               - \frac{2i}{\ik}\;\partial \psi^6  
$$
where $J^\mu=i\partial X^\mu,\,\psi^\mu$ for $\mu=0,\ldots,5$ are the 
free fields from the $\R^{5,1}$ subtheory and where $J^6 = i\partial \phi$ 
is formed from the Feigin-Fuchs boson in the linear dilaton theory, with 
associated free fermion $\psi^6$. An explicit expression for $G^7$ in the 
AdS$_3 \times T^4$ model can be found e.g.\ in \cite{EGP}; it is rather 
similar to the SU(2) supercurrent because of the relation of AdS$_3$ 
and the SU(1,1) WZW model. 

\smallskip\noindent
We are interested in BPS branes here, so in particular the world-sheet 
supercurrent must be preserved by the corresponding boundary conditions, 
i.e.\ we must have $G(z) = \pm \overline{G}(\bar z)$ along the boundary 
of the world-sheet. It is easy to see that, in the 
NS 5-brane case, this enforces Neumann conditions on the linear dilaton: 
All the branes discussed here will end on the stack of NS 5-branes. 
As usual, the allowed number of Neumann directions in the remaining 
target dimensions depends on whether we work in IIA or IIB string
theory.
\newline\noindent
It is convenient to pass to a new set of SU(2) currents defined by 
$$
\tilde J^a := J^a + \frac{i}{\ik} \; f^a_{\ b c}\, \psi^b \psi^c \ \ ,
$$
which satisfy the WZW operator product expansion (\ref{JOPE1}) at level 
$\ik -2$ instead of $\ik$, and which commute with the fermions $\psi^a$. 
(On AdS$_3$, one proceeds analogously for the currents of the 
SU(1,1) subtheory.) Using the new fields, it becomes clear that the 
CFTs under consideration split into tensor products of a bosonic 
part with ten free fermion, and that the same is true for the boundary 
states describing branes in the NS 5-brane or  AdS$_3 \times S^3 \times T^4$
superstring background. As a consequence, space-time supersymmetry 
charges can be built up from the usual spin fields, and also the 
behavior under the GSO projection carries over from the flat target 
case. In particular, the partition functions of our supersymmetric 
branes in $M^7 \times S^3$ vanish, meaning that the spectra are 
tachyon-free and stable. Geometric arguments for stability of branes 
in $S^3$, based on the non-vanishing B-field and on the quantization 
condition on the brane labels $\alpha$, were presented recently in 
\cite{BaDoSc,Paw}. Our analysis of the equations of motion will, 
among other things, confirm these results. 

\smallskip\noindent
Let us now collect formulas for the physical vertex operators 
necessary to compute the low-energy effective action in the 
supersymmetric case. Those corresponding to gauge bosons and scalars 
on the external part of the brane world-volume can be treated as usual, 
therefore we concentrate on the 'Higgs scalars' associated with 
the $S^3$-directions. In the picture with superghost number -1, 
their vertex operators are 
$V^{(-1)}_H [\tA](x) = V[e_k] :\!\psi^aV[\tA_a]\!:(x)$ 
with corresponding states 
$$
|\, V^{(-1)}_H [\tA]\,\rangle = 
a^{Jm}_a\;|k\rangle\otimes \psi^a_{-\frac{1}{2}} |Y^J_m\rangle\ \ .
$$ 
Here, the SU(2) highest weight states are tacitly understood to contain 
the vacuum of the fermionic Fock space. Moreover, we now approximate 
the external part $M^7$ by flat space-time such that $k$ is a 
$\tilde p$-momentum where $\tilde p$ is the number of Neumann 
directions in $\R^7$. As before, the parameters $\tA_a$ are subject 
to the physical state condition $\tL^a\tA_a = 0$. 
\newline\noindent
For the computation of 3- and 4-point functions, we also need vertex 
operators in the ghost number 0 picture. From the usual commutation 
relations between $G_{\rm SWZW}$ and $\psi^a,\,J^a$, one finds the SU(2) 
component of the corresponding states to be  
\ba
|\,V^{(0)}_H [\tA]\,\rangle  
 & = & G_{-\frac{1}{2}} |\, V^{(-1)}_H [\tA]\,\rangle 
\nn\\
& = &  a_a^{J m}\; \Bigl(\, \tilde J^a_{-1} 
  - \frac{2}{\ik}\; \delta_{c b}\; \tilde J^c_{0}\,
                \psi^a_{-\frac{1}{2}}\,\psi^b_{-\frac{1}{2}}      
 - \frac{i}{\ik}\; f^a_{\ b c}\; \psi^b_{-\frac{1}{2}}\,\psi^c_{-\frac{1}{2}}
     \, \Bigr)\;|Y^J_m\rangle\ \ .
\nn\ea
It is now straightforward to compute the 3-point function 
$$
\langle\;  V^{(-1)}_H [\tA_1](x_1) \;V^{(-1)}_H [\tA_2] (x_2)\; 
V^{(0)}_H [\tA_3] (x_3)\;\rangle\ \ ,
$$
and the result is given by the terms displayed in eq.\ (\ref{3pt}). 
In the supersymmetric case, there are no  higher order corrections in 
$\ap$. Likewise, the leading contribution to the 4-point function is 
as in the bosonic WZW model. 

\noindent All in all, the low-energy effective action (\ref{geomeffact}), 
supplemented by the non-abelian Yang-Mills-Higgs theory for the external 
degrees of freedom, also describes the dynamics of supersymmetric branes 
in $\R^7 \times S^3$. Of course, there are additional terms we did not 
discuss, which mix the fields from the external gauge theory with 
the Higgs scalars from the $S^3$ factor. To leading order in $\ap$, 
they can be copied from the effective action for flat branes in 
$\R^{10}$. We will not discuss space-time fermions here, whose action 
could in principle be computed from world-sheet Ramond fields, or by 
exploiting space-time supersymmetry.

\section{Classical solutions and bound states} 

Having found the action that describes the dynamics of branes
on SU(2), we can start analyzing the classical equations of 
motions and their solutions.  The equations of motion which derive 
from the action (\ref{geomeffact}) read 
\be
\label{action} 
\tL^a\, \tF_{a b} \, + \,  [\, \tA^a \astk \tF_{a b}\, ] \ = \ 0 \ \ .
\ee 
Geometrically, this means that the curvature of the solutions 
must be covariantly constant. 
When we deal with a stack of $N$ branes which wrap the fuzzy sphere 
labeled $\alpha$, the equations are to be read as equations in 
$\Mat(N) \otimes \Mat (2\a+1)$, and $\tL^a = 1 \otimes \tL^a$ 
acts trivially on the Chan-Paton factors.

\noindent 
We will first concentrate on constant solutions and then extend our 
discussion to non-constant ones towards the end of the section. 
On constant fields $\tS_a$, i.e.\ those satisfying $L^a \tS_b = 0$ 
for all $a,b$, the equations of motion (\ref{action}) simplify to 
\be 
\label{caction} 
\Bigl[\, \tS^a\astk \;\, [\, \tS_a \astk \tS_b\, ] \, - \, i\, 
 \tf_{a b c}\, \tS^c \,\Bigr] 
\ = \ 0 \ \ . 
\ee 
As we will see in a moment, even the constant solutions give rise to 
a rather non-trivial pattern of D-brane configurations. The equations 
(\ref{caction}) possess two different types of solutions. Obviously, 
they are solved by any set of commuting matrices $\tS_a \in 
\Mat(N) \otimes \one_{2\alpha+1}$. These solutions correspond to 
marginal fields in the SU(2) theory which translate our $N$ D-branes 
in the group target, but they have no influence on the radius. The new 
positions of the branes are determined by the eigenvalues of the matrices 
$\tS_a$. If some of these eigenvalues agree, smaller stacks of spherical 
branes appear in the final configuration. All these statements are familiar 
from D-branes in flat space \cite{Wit} and they can be verified rigorously 
along the lines of \cite{RS2}. 

\smallskip\noindent
Remarkably, there exists a second type of constant solutions without 
any flat space analogue. They are given by constant flat connections, 
or equivalently, by matrices $\tS_a \in \Mat(N)\otimes \one_{2\alpha+1}$ 
satisfying the commutation relations  
\be \label{nsol}
 [\, \tS_a,\tS_b\, ] \ = \ i \, \tf_{ab}^{\ \ c}\; \tS_c 
\ee
which, on constant solutions, are equivalent to 
$\tF_{a b} (\tS) =  0$. Equations (\ref{nsol}) simply 
characterize $N$-dimensional representations of su(2). To any 
such representation $\pi_N$ we can assign a partition 
$(n_i)_{i=1,\dots,r}$ of 
$N = n_1 + \dots + n_r$ by choosing $n_i$ to be the dimensions of the 
irreducible sub-representations in $\pi_N$. This partition characterizes 
the original 
representation uniquely up to equivalence, hence it provides a
simple classification for solutions of the second type. Our objective 
is to identify these solutions with known configurations of branes. 
Rigorous statements about this identification are difficult since 
the boundary perturbations generated by the set $\tS_a$ are 
marginally relevant. This implies that there is no continuous 
trajectory in the moduli space of conformal field theories that 
we could follow. In this respect, the same problems arise in the 
study of marginally relevant operators as with open string tachyons and 
their condensation \cite{ostach1,Sen,RS2,ostach2}. Nevertheless, we 
will provide strong evidence for the pattern of renormalization group 
fixed points we are about to suggest. 

\smallskip\noindent   
To this end, recall that the state space of our boundary theory 
is given by $\Mat(N) \o \cH_\alpha$ where $\cH_\alpha$ is the state
space of a single brane with label $\alpha$, and $\Mat(N)$ are the 
wave functions associated with the Chan-Paton factors. The modes 
$j^a_n$ of the chiral fields $j^a(z)$ generate an action of $\widehat
{\rm SU}(2)_\ik$ on the second tensor factor. Moreover, there is an action 
of su(2) coming with our solution $\tS_a$. It is defined through 
the commutator of $\tS_a$ with the Chan-Paton matrices in the first 
tensor factor $\Mat(N)$ of our state space, i.e.\ $\tS_a$ act as 
$ad(\tS_a)$. Once we turn on the perturbation 
$$ S_{\rm pert} \ = \ \int dx\; :\! j^a V[\tS_a]\!:(x) \ =\  
   \int dx\; j^a (x) \tS_a\ \ , $$ 
these two `symmetries' 
are broken because the perturbing operator couples the SU(2) current
to the quantum mechanical su(2) of the Chan-Paton factors. From 
experience with similar couplings of spin and orbital angular momentum 
in quantum mechanics, one may suspect that only the sum $j^a + \tS^a$ 
has a good chance to be realized as a symmetry in the perturbed theory.
Indeed, the fact that our matrices $\tS_a$ obey the equations of motion 
(\ref{nsol}) implies that the sum $j^a_n + ad(\tS^a)$ gives rise to an 
action of the current algebra $\widehat{\rm SU}(2)_\ik$ with level $\ik$ 
on the state space $\Mat(N) \o \cH_\alpha$. 
Hence, if we assume that the action survives the perturbation, we can 
use it to determine the representations of $\widehat{\rm SU}(2)_\ik$ 
that are present in the 
perturbed theory. The decomposition is easy to find with the 
help of zero modes, and it turns out that there is a unique way 
of combining their characters back into partition functions of 
the known D-brane theories. In principle, the partition functions
do not quite suffice to recover all information about the D-branes since
they are invariant under translations on the background. However, 
the position of the branes can be detected by bulk fields \cite{FFFS}, 
which are sensitive only to the scalar part of the Chan-Paton factors. 
The fact that $\tr \tS_a$ vanishes for all solutions of eqs.\ (\ref{nsol}),
shows that, to first order in the perturbative expansion, the branes are 
not translated, i.e.\ their classical counterparts are conjugacy classes 
centered around the origin $e$, as listed after eq.\ (\ref{Theta}). 

\smallskip\noindent
The above analysis leads to the following scenario: When we start from 
a stack of $N$ $\a$-branes centered around $e$, the solution $\tS_a$ 
of eqs.\ (\ref{nsol}) corresponds to the fixed point of an RG flow 
that takes us to a new brane configuration consisting of a superposition 
of spherical branes, again centered around $e$. These branes can have 
different radii, and they can be stacked or just single branes. To be 
more precise, we use the associated partition $(n_i)$ and introduce 
spins $J_i = (n_i-1)/2$. Then the boundary state for the final D-brane 
configuration can be written as sum of 'elementary' boundary states 
$|\beta\rangle$ for single spherical branes,   
\be \label{multip}
|(\tS_a)\rangle = \sum_\beta\; N_\beta\,|\beta\rangle 
\quad\ {\rm with}\ \ N_\beta = \sum_i N_{\alpha J_i}^\beta\ \ .
\ee
Here, $N_{\alpha J}^\beta$ denote the fusion rules of su(2). The
multiplicity $N_\beta$ is the size of the stack of branes of type 
$\beta$ contained in the final configuration. 
\medskip

\noindent
As an example, let us consider an initial stack of $N$ D0-branes at the
origin and choose an irreducible $N$-dimensional representation 
$\tS_a$ of su(2). Then, according to our rules, this will decay 
into a single brane of type $\beta = (N-1)/2$. In other words, the 
stack of D0-branes has evolved into a single brane that wraps a 
non-degenerate 2-dimensional (fuzzy) sphere. One could say that, in
this way, we have dynamically created two new dimensions on the 
expense of breaking the gauge symmetry from U($N$) to U(1). 
Had we used some reducible representation $\tS_a$ instead, with 
partition $(n_i)_{i= 1,\dots,r}$, stacks of $N_{\beta} = \# \{\, n_i\, 
|\, n_i = 2\beta + 1\}$ branes would have appeared, which wrap the 
spheres with labels $\beta$. In case there were degeneracies among 
the $n_i$, the gauge group would be only partially 
broken to $\bigtimes_\beta\,$U($N_\beta$). 

\smallskip\noindent
We want to use the example of $N$ D0-branes blowing up into 
a single D2-brane to provide more evidence for our identification of 
the fixed points. The idea is to exploit that we certainly know the 
tension of all the maximally symmetric branes. If our identification 
is correct, the leading term in the difference of tensions between 
the initial and the final brane configuration has to agree with the 
effective action $\cS(\tS_c)$ evaluated at the classical solution. 
The strategy is similar to the one used in \cite{ReRoSc}. 

\noindent
The tension of a single brane is essentially given by the logarithm 
of the so-called $g$-factor of the boundary conformal field theory, 
see e.g.\ \cite{HarMoo,ReSh1}. The latter is defined as the 1-point function 
of the identity operator \cite{AffLud}, and for a brane of type $\a$ 
in the SU(2) WZW model it is given by \cite{Car} 
$$ 
    g_{\alpha} \ = \ \frac{S_{\alpha\, 0}}{\sqrt{S_{0\, 0}}} \ = \ 
  \ = \ \left(\frac{2}{\ik+2}\right)^\frac{1}{4} \ 
      \frac{\sin \frac{(2\a+1)\pi}{\ik+2}}{
      \sin^{\frac{1}{2}}\frac{\pi}{\ik+2}} \ \ . 
$$ 
Here $S_{IJ}$ denotes the modular S-matrix of the WZW model. 
For the initial stack of $N$ identical D0-branes, the $g$-factor is 
$g_{N\, {\rm D0}} = N \, g_{0}$. If we expand the difference of the 
logarithms in a power series around
$1/\ik =0$ we find to leading order
$$ \ln \frac{g_{(N-1)/2}}{g_{N\, {\rm D0}}} \ = \ 
   - \frac{\pi^2}{6}\, \frac{N^2-1}{\ik^2}\ \ . 
$$ 
The minus sign shows that the tension of the final state is lower
than the one of the original stack of branes,  in agreement 
with the `$g$-theorem' stating that the $g$-factor should always 
decrease along the renormalization group trajectory \cite{AffLud}. 

\noindent
Now we have to compare our result on the $g$-factors with the value of the 
action $\cS(\tS_a)$ where $\tS_a$ is a solution of eqs.\ (\ref{nsol}) 
corresponding to an irreducible $N$-dimensional representation of 
su(2). Because $\tS_a$ is constant, all the terms in which $\tL_b$ 
acts on $\tS_a$ vanish, and after a short computation we obtain 
$$    \cS(\tS_a) \ = \ - \, \frac{(2\pi\ap)^2}{6 \a' \ik} \ \tr 
      (\, \tS_c \ast \tS^c\, ) \ \ . 
$$ 
Up to normalization, the term in the argument of the trace is 
simply the Casimir element of su(2) evaluated in the irreducible 
representation of spin $\a = (N-1)/2$, i.e.\ the argument of the 
trace is proportional to $\a (\a+1) = (N+1)(N-1)/4$. Taking 
the normalization into account, we find 
$$   \cS(\tS_a) \ = 
  \ln \frac{g_{(N-1)/2}}{g_{N\, {\rm D0}}} \ \ 
$$ 
as desired. 
Note that the value of the action (the potential energy of the Higgs 
fields) at the new fixed point is actually lower than in the original 
state which means that the $N$ D0-branes have formed a bound state. 

\medskip\noindent
So far we have only discussed constant solutions to the full
equations of motion (\ref{action}). For D0-branes, this
discussion is complete since any field $\tA_a$ on a single 
point is constant. On the other hand, we have obtained all 
other branes $\alpha$ as bound states of D0-branes. Combining 
the two facts, it should be possible to understand non-constant 
classical solutions, as well. To this end, we first rewrite the 
equation of motion (\ref{action}) according to  
\be \label{eom} 
\Bigl[\; \tB^a \astk \; [\, \tB_a \ast \tB_b\, ] \, -\, i\, 
    \tf_{a b c}\, \tB^c \, \Bigr] \ = \ 0  
\ee
with $\tB_a = \tY_a + \tA_a$ as before. Now the general 
equations of motion (\ref{action}) have become formally identical 
to (\ref{caction}) which govern the dynamics of D0-branes, 
and solving them amounts to finding matrices 
\hbox{$\tB_a \in \Mat(N)\o \Mat(2\a+1)$} that satisfy our 
familiar eqs.\ (\ref{eom}). From any such solution 
$\tB_a$ we can then obtain a field $\tA_a = \tB_a - \tY_a$ that 
will carry us into a new RG fixed point upon perturbation. Hence, it 
only remains to identify this fixed point. In case that $\tB_a$ is 
a representation of su(2), we claim that the fixed point is determined 
by the partition $(n_i)$ characterizing the representation $\tB_a$ of 
su(2) along with the rules we have used for stacks of D0 branes above. 
This means that there will appear $N_{\beta} = \# \{\, n_i\,|\, n_i = 
2\beta + 1\}$ branes wrapping the spheres with labels $\beta$ in the 
final configuration. 

\noindent Note that this pattern is consistent with the one proposed 
in (\ref{multip}). In fact, if $\tA_a = \tS_a$ is constant, i.e.\ if 
it commutes with $\tY_a$, then the elements $\tB_a$ generate the 
tensor product of the $(2\a+1)$-dimensional irreducible representation 
with the representation $\tS_a$. Hence, the partition associated with 
$\tB_a$ can be obtained from the one that comes with $\tA_a$ upon fusing 
with the spin $\a$ representation.

\smallskip\noindent  
Our claim on the nature of the fixed points described by  
su(2) representations $\tB_a$ can be substantiated with the  
same methods as before. In particular, one can again compare  
the $g$-factors of the proposed fixed points with the value  
of the effective action. When evaluated on a solution of our  
equation of motion (\ref{eom}), the latter becomes  
\be\label{actonsols} 
\cS(\tB_a) = \frac1{12}\; \ \tr \left(\,  
 [\, \tB_a \ast \tB_b\, ]  \ast  [\, \tB^a \ast \tB^b\, ]   
 + \frac{2}{\ap\ik}\;\tY_a \ast \tY^a\, \right)\ \ .
\ee 
The formula shows that commuting matrices $\tB_a$ cause 
the action to increase, i.e.\ they lead to a final 
configuration which has a larger mass than the one we 
started with. On the other hand, 
the lowest action among all the solutions we have described 
is assumed for $\tB_a$ being an irreducible representation 
of su(2). This corresponds to the stack of branes merging into  
a single brane with $\beta = N \a + (N-1)/2$.  

\noindent
In addition, one may compute the second variation of the action
and evaluate it at an $N(2\a+1)$-dimensional su(2) 
representation $\tB_a$ to get
$$\delta^2 \cS \ = \  \tr \Bigl( - [\, \tB_a \astk  
\delta \tB_b\, ]\, \ast \,  [\, \tB^a\astk \delta \tB^b\, ] 
 + [\, \tB_a\astk \delta \tB^a\, ] \, [\, \tB_b\astk 
  \delta \tB^b\, ]\, \Bigr)\ \ . 
$$
Using the fact that the map $\tA \mapsto [\tB_a,[\tB^a,\tA]]$ 
is invertible on all traceless matrices $\tA$, one can easily 
see that the su(2) representations $\tB_a$ represent a local 
minimum of the action (except from the obvious zero-modes 
associated with rigid translations), in agreement with the 
results of \cite{BaDoSc,Paw} on the stability of the
spherical branes in WZW models.

\smallskip\noindent 
The previous analysis of non-constant solutions may be turned around  
and used to `derive' the equations of motion (\ref{action}) for an  
arbitrary brane of type $\alpha$ from those of D0-branes.$\,$\footnote{This 
argument was suggested to us by J.\ Maldacena.} In fact, we may think 
of  the fields $\tA_a$ on the $\a$-brane as fluctuations around the 
constant solution $\tS_a$ which takes us from $N$ D0-branes to 
the single D2-brane. Now we recall that $\tS_a$ is an irreducible 
$(2\a+1)$-dimensional representation of su(2). The world-volume 
algebra of the $\a$-brane, on the other hand, contains the 
elements $\tY_a$ which also generate an irreducible su(2) 
representation of dimension $2\a+1$. Since any two irreducible su(2) 
representation of the same dimension are equivalent, we can identify 
$\tS_a$ with $\tY_a$ so that the idea of studying small 
fluctuations around $\tS_a$ motivates to insert the combination 
$\tY_a + \tA_a$ into the equations (\ref{caction}). The resulting 
equations are given by (\ref{eom}) and do indeed describe 
the full dynamics of the $\a$-brane.  

\smallskip\noindent
Let us finally note that our discussion of the constant solutions 
coincides with the  `absorption of boundary spin' that has appeared in 
the work of Affleck and Ludwig on the multichannel Kondo-problem 
\cite{AffLud}. In that context, an irreducible representation $\tS_a \in 
\Mat(N)$ is reinterpreted as the quantum mechanical spin variable of a 
spin $J = (N-1)/2$ impurity. According to Affleck and Ludwig, this 
system evolves into a new low-temperature fixed point in which the 
boundary spin gets absorbed. 
\medskip

\section{Conclusions and outlook} 

\noindent 
In this paper we have studied the dynamics of branes in $M^7\times S^3$ 
which, classically, wrap some $S^2 \subset S^3$, focussing on the 
limit where the radius $R \sim \sqrt{\a'\ik}$ of $S^3$ grows very 
large. The fact that the spherical conjugacy classes become fuzzy 
spheres in the quantum theory has a number of 
consequences for the low-energy effective action: 

\noindent  First, the gauge fields on the external 
part of the branes become non-abelian already on a single brane. 
The size of the gauge group U$(N\,(2\alpha+1))$ depends on the 
number $N$ of coinciding branes and on the label $\alpha$ of the 
fuzzy sphere that is wrapped in $S^3$. 

\noindent Secondly, the Higgs potential for the scalars associated 
with the internal $S^3$ consists of a non-commutative Yang-Mills 
theory on the fuzzy sphere plus a non-commutative analogue of the 
Chern-Simons action. As expected on general grounds, this LEEA 
enjoys invariance under non-commutative gauge transformations. 

\smallskip\noindent We have studied solutions to the equations of 
motion derived from the non-com\-mut\-ative part of the effective 
action. We found that a certain class of constant field 
configurations, namely those obtained from pairwise commuting 
constant gauge fields, are associated with moduli that 
trigger rigid translations of branes in the $S^3$ background. The 
more interesting type of solutions is given by reducible or irreducible 
representations of su(2) and has no analogue for flat branes in 
$\R^d$. These solutions describe fixed points of RG flows driven by 
marginally relevant perturbations of the world-sheet theory. At the 
fixed points, the initial stack of identical branes has evolved into a 
bound state of SU(2) branes, whose precise form is determined by 
the su(2) representation. In contrast to the 'constant commuting 
case', the new type of solutions given in (\ref{nsol}) can change the 
radii of the branes. In this way, we can in particular interpret
2-dimensional spherical branes as bound states formed from a stack 
of D0-branes.  

\noindent
As we pointed out above, gauge theories on fuzzy spheres 
were discussed in the literature before \cite{Mad,Wat,FFT,Madbook,Klim},
and the field equations that were derived from such actions are very 
similar to the equations we have studied here. The non-commuting 
solutions (\ref{nsol}) are a common feature of these cases, but 
the model we have derived here from perturbative string theory 
is distinguished by the property that it also admits the 
commuting solutions corresponding to rigid movements in $S^3$.

\smallskip\noindent 
Based on what we have described above, one cannot exclude the 
existence of additional solutions which could lead the system 
into new types of bound states. Even if such configurations 
exist, it is likely that the single spherical brane realizes 
the global minimum of the action. We have not been able to prove
this in the general case but at least it is true for $N (2\a+1) 
= 2$.  

\smallskip\noindent
Extending the present analysis to finite radius $R \sim \sqrt{\a'\ik}$ 
is another challenge. From the results of \cite{AlReSc} one may 
be tempted to believe that part of the picture could survive if the Lie 
algebra su(2) is replaced by the corresponding quantum group and
its representation theory. The relation with quantum group 
representations was also stressed in \cite{FeLeSa}; their importance 
also transpires from the general investigations in \cite{FFFS,FFFStop,BPPZ}.  
Note that in the Kondo problem, the level $\ik$ counts the number of 
conduction bands. Hence, Kondo physics explores the region of small level. 
In this context one may recall that for  $N = \ik n/2$ and 
$n$ integer, which of course requires to go beyond a perturbative 
analysis in $1/\ik$, the known fixed-points are described by free rather 
than interacting theories \cite{AffLudUS}. This suggests that new phenomena 
occur for stacks of D0-branes on a sphere of finite size. For instance, the 
analysis of Affleck and Ludwig \cite{AffLudUS} shows that a stack of 
$N$ D0-branes can evolve into a single spherical brane with label 
$\alpha$ where $\a \leq \ik/2$ is determined from $N$ by the rule $N = 
\ik n/2 + \alpha$. In view of the work \cite{Li}, 
which compares 'giant gravitons' with spherical branes on SU(2), such 
finite level effects should be related to the 'stringy exclusion principle'.  

\smallskip\noindent
As we have pointed out before, most of our statements admit obvious
generalizations to other group manifolds. The necessary CFT ingredients 
can be recovered e.g.\ from \cite{FFFS,BaDoSc}. In view of the 
perturbative treatment of \cite{Vol} and of the general definition of 
non-commutative world-volumes from boundary CFT given in \cite{AlReSc}, 
we believe that the lessons we have learned are much more universal. 
Flows from stacks of D0-branes to branes whose classical counterparts 
have higher-dimensional world-volume should be a common phenomenon in  
string compactifications. It is a distinguished feature of string theory 
that the world-sheet perspective provides natural descriptions of effects 
that appear like a `topology change' in the space-time picture, 
e.g.\ the dynamical generation of extra world-volume dimensions as 
discussed above, or the rotation of Dirichlet into Neumann conditions 
for free bosons by marginal operators \cite{RS2}. We hope that 
extensions of our results may also provide some quantitative insight 
into tachyon condensation and bound state formation on other more 
complicated string backgrounds which possess an exact conformal field 
theory description. This applies especially to Gepner models and their 
associated D-brane theories \cite{ReSh1,Gepbran}. Another step in this 
direction was taken recently in \cite{ReRoSc}, and we hope to return 
to these issues in the future. 
\medskip

\noindent 
\acknowledgments{ 
We would like to thank I.\ Brunner, M.\ Douglas, S.\ Fredenhagen, 
J.\ Fr\"ohlich, K.\ \gaw, H.\ Grosse, J.\ Louis, D.\ L\"ust, J.\ Madore, 
J.\ Maldacena, G.\ Reiter, S.-J.\ Rey, M.\ Rozali, J.\ Teschner and 
S.\ Theisen for 
useful and stimulating discussions. This work was started during a stay 
of the three authors at the ESI for Mathematical Physics in Vienna.}

\end{document}